\documentclass{raa_twocolumn}            

\usepackage{graphicx,times}             
\usepackage{natbib}
\usepackage{amssymb,amsmath}
\bibpunct{(}{)}{;}{a}{}{,}
\usepackage[pagebackref=true]{hyperref}

\begin{document}

  \title{Design and in-orbit calibration of the MXT optics}

   \volnopage{Vol.0 (202x) No.0, 000--000}      
   \setcounter{page}{1}          

   \author{C. Feldman 
      \inst{1}
   \and J. Pearson
      \inst{1}
   \and G. Butcher
      \inst{1}
   \and R. Willingale
      \inst{2}
   \and P. O'Brien
      \inst{2}
   \and J. P. Osborne
      \inst{2}
   \and K. Mercier
      \inst{2}
   \and JM. Le Duigou
      \inst{2}
   \and D. G$\ddot{o}$tz
      \inst{3}
   }

\institute{Space Park Leicester, University of Leicester, United Kingdom; {\it chf7@le.ac.uk}
\and
School of Physics and Astronomy, University of Leicester, United Kingdom;
\and
Ctr National d'Études Spatiales, Toulouse, France;
\and
AIM-CEA/DRF/Irfu/Departement d’Astrophysique, Universite Paris-Saclay, France;}


\abstract{The Microchannel X-ray Telescope (MXT) is one of four instruments on the Space-based multi-band astronomical Variable Objects Monitor (SVOM) satellite mission, launched on the 22$^\mathrm{nd}$ June 2024. The MXT is a narrow-field-optimised, lobster eye X-ray focusing telescope, consisting of an array of 25 square MPOs, with a focal length of 1.14 m and working in the energy band 0.2 - 10 keV. The design of the MXT optic (MOP) is optimised to give a 1$^{\circ}$ FoV to match the detector size, but the optic has the unique characteristics of a lobster eye design, with a wide FoV of 6$^{\circ}$ diameter, and a PSF, which is constant over the entire FoV. The MPOs on the Flight Module (FM) MOP have a pore size of 40 ${\mu}$m giving the optimum thicknesses across the aperture of 2.4 mm in the centre and 1.2 mm at the edges. Using specific target sources, the in-orbit calibration of the optic is here described, and compared to the extensive on-ground calibration, which was carried out at the PANTER test facility, MPE, Germany. The design and limitations of the electron diverter, situated directly behind the optic, are also discussed.}

   \authorrunning{C. Feldman}            
   \titlerunning{In-orbit calibration of the MXT optics}  

   \maketitle

%
%
\section{Introduction}           
\label{sect:intro}
\subsection{Lobster eye optics}
\label{lob}
The lobster-eye geometry for X-ray imaging, first proposed by Angel in 1979(\cite{ang}), has since been widely developed using Microchannel Plates (MCPs) or Micropore Optics (MPOs) (e.g. \cite{theory, wilks, fraspie, kaa}). Compared to traditional Wolter-type optics(\cite{wolt}); this design enables a significantly wider field of view with substantially lower mass.

As detailed in \cite{hudec_2022}, large-area lobster-eye telescopes can be constructed by tessellating spherically slumped MPO tiles onto a precision machined, spherical support frame. Each MPO consists of square, ~40 $\mu$m pores etched through a glass substrate and subsequently slumped to match the spherical curvature. When mounted on a frame with the same spherical radius, the optic channels align toward the common center of curvature, enabling wide-field X-ray focusing. Typically, a thin aluminium film is applied (70 - 120 nm in thickness) over the front pore apertures, which provides both a filter for optical and UV light, and a good thermal surface. 

\begin{figure}
	\centering
		\includegraphics[width=0.31\textwidth]{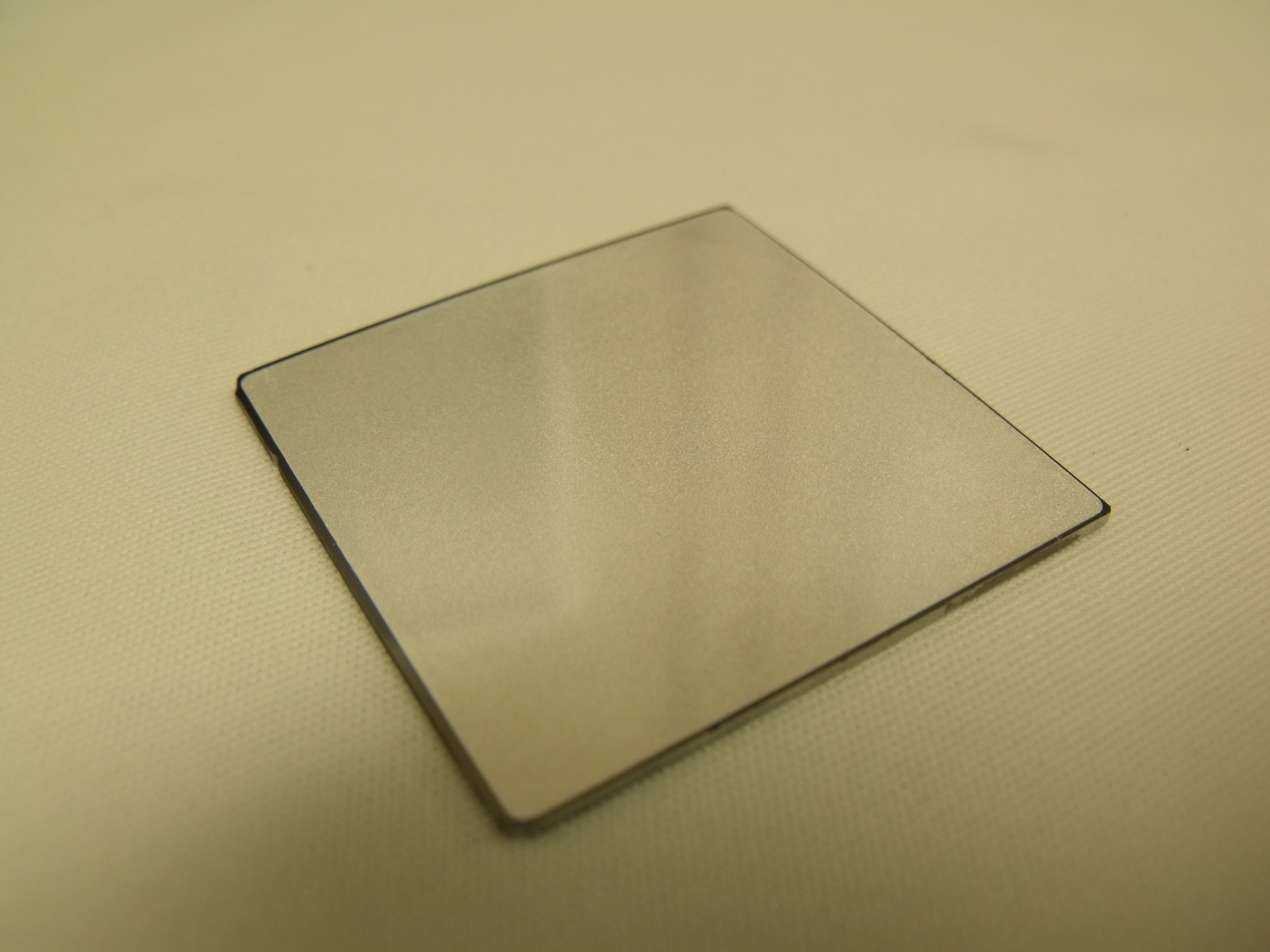}
		\includegraphics[width=0.31\textwidth]{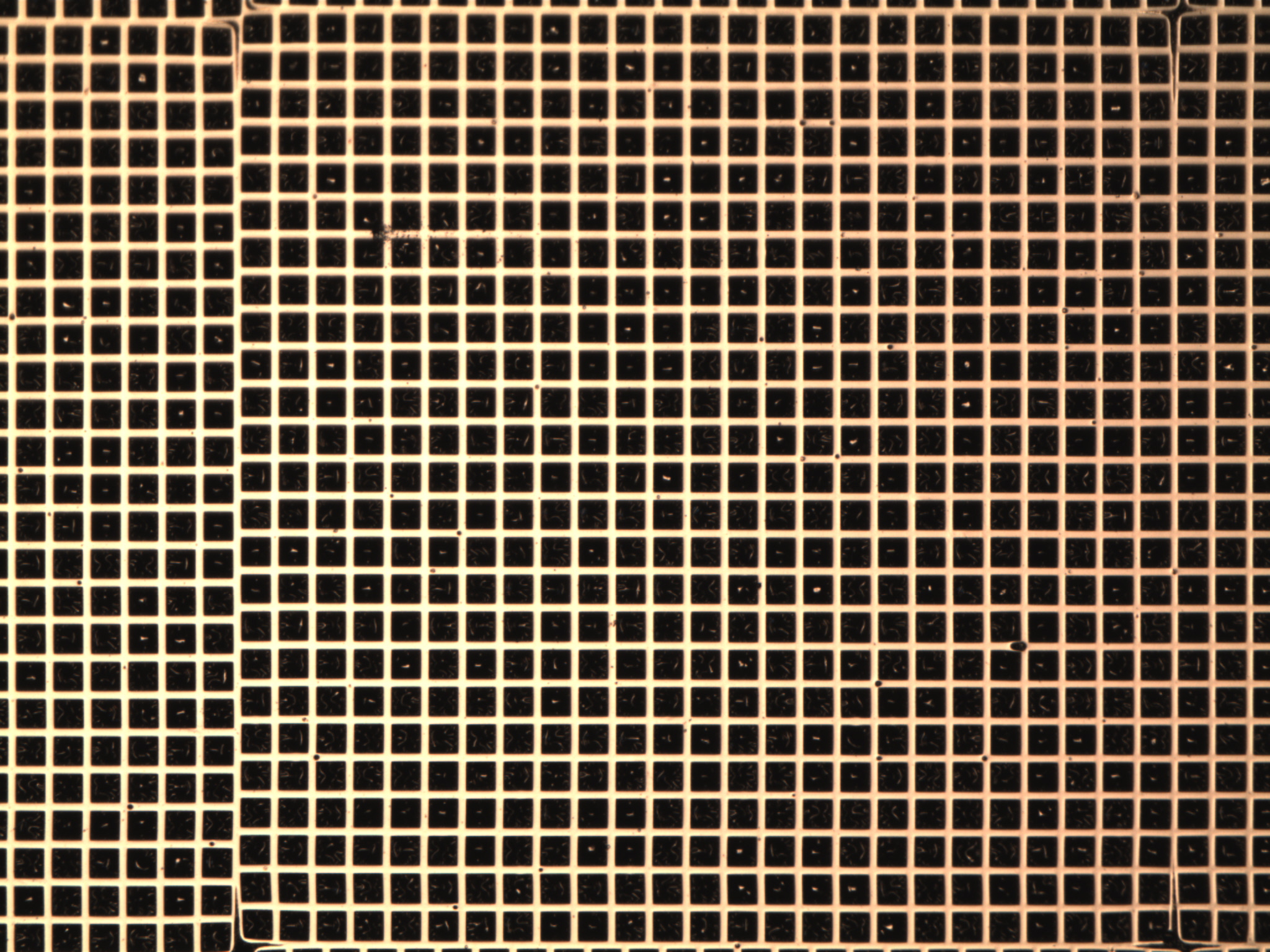}
			\caption{\textit{Top}: a single, 40 mm by 40 mm square, iridium coated, aluminium filmed, 1.05 mm thick MPO Produced by Photonis France SAS. \textit{Bottom}: a microscope image of a series of individual 40 $\mu$m pores on an MPO.}
	\label{fig2}
\end{figure}

The geometry of each MPO comprises a square packed array of microscopic pores, each with a square cross-section. A description of the production of the MPOs can be found in \cite{thspie} and \cite{photmpo}. An example of an MPO and a microscope image of the pores is shown in Figure \ref{fig2}.

\begin{figure}
	\centering
		\includegraphics[width=0.45\textwidth]{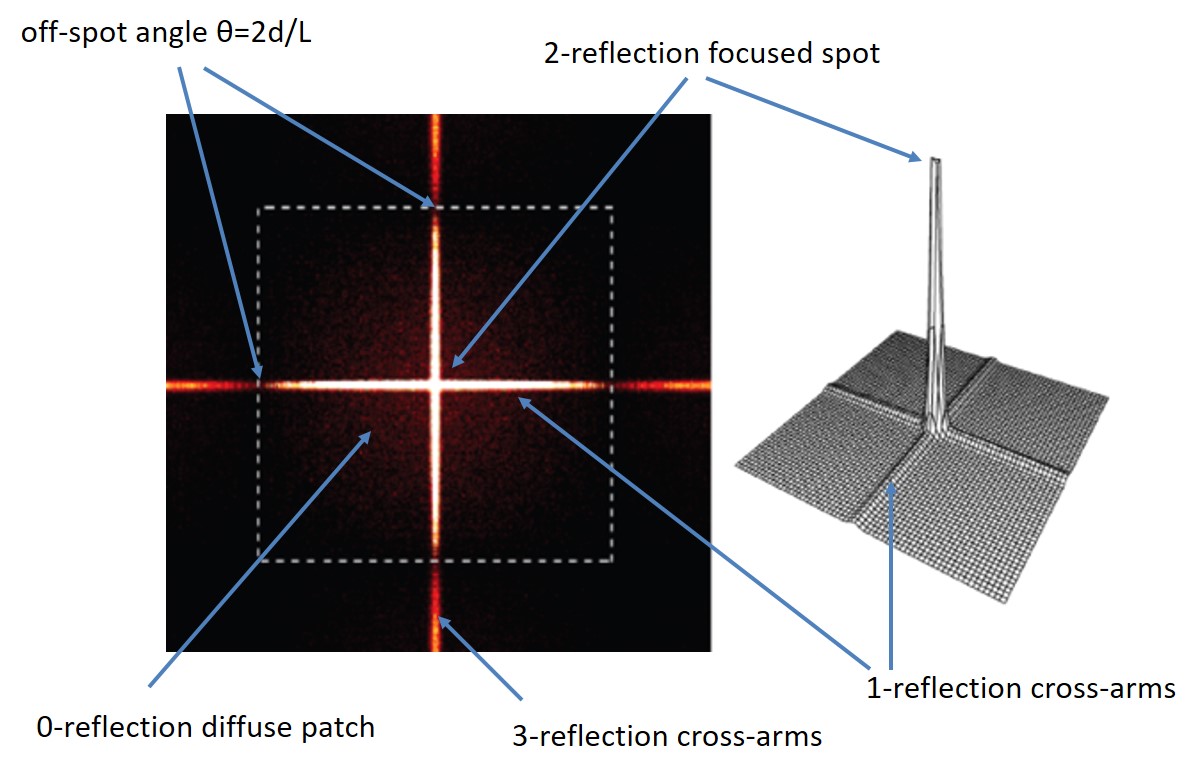}
			\caption{The distinctive PSF created by a slumped MPO(\cite{dickspie}).}
	\label{geom2}
\end{figure}

The point spread function (PSF) of a single MPO exhibits a characteristic cross-shaped structure. The central focused spot results from photons undergoing two successive grazing-incidence reflections off adjacent pore walls. The orthogonal cross-arms arise from rays that undergo an odd number of reflections—typically single reflections—before reaching the focal plane. A diffuse background is produced by rays that either pass straight through the pores (zero reflections) or experience multiple, non-focusing reflections. Figure \ref{geom2} illustrates the PSF from a single MPO.

\subsection{The MXT optic}
\label{MXT}
The Space-based multi-band astronomical Variable Objects Monitor (SVOM) mission includes four onboard instruments (Cordier et al. this issue), supported by dedicated ground-based facilities. Among the on-board instruments, is the French led Microchannel X-ray Telescope (MXT) operating in the 0.2–10 keV energy range, providing high-sensitivity observations of gamma-ray burst (GRB) afterglows with rapid and precise localization.

The initial GRB positions are provided by another on-board instrument, ECLAIRs (G\"{o}tz et al. this issue, Godet et al. this issue), an X-/Gamma-ray wide-field (89$^{o}$ x 89$^{o}$) coded-mask camera working in the 4 – 250 keV energy range, which the MXT then refines and enables detailed spectral and temporal analysis of the afterglow emission.

\subsection{Design}
\label{sect:design}
The design of the MXT optic (MOP) is described in \cite{svomspie22} and is narrow field optimised to give a $\sim$1$^{o}$ detector limited field of view, but the optic has the unique characteristics of a lobster eye design, with a wide field of view $>$ 6$^{o}$ diameter, and a PSF which is constant over the entire field of view of the optic.

\begin{figure}
	\centering
		\includegraphics[width=0.45\textwidth]{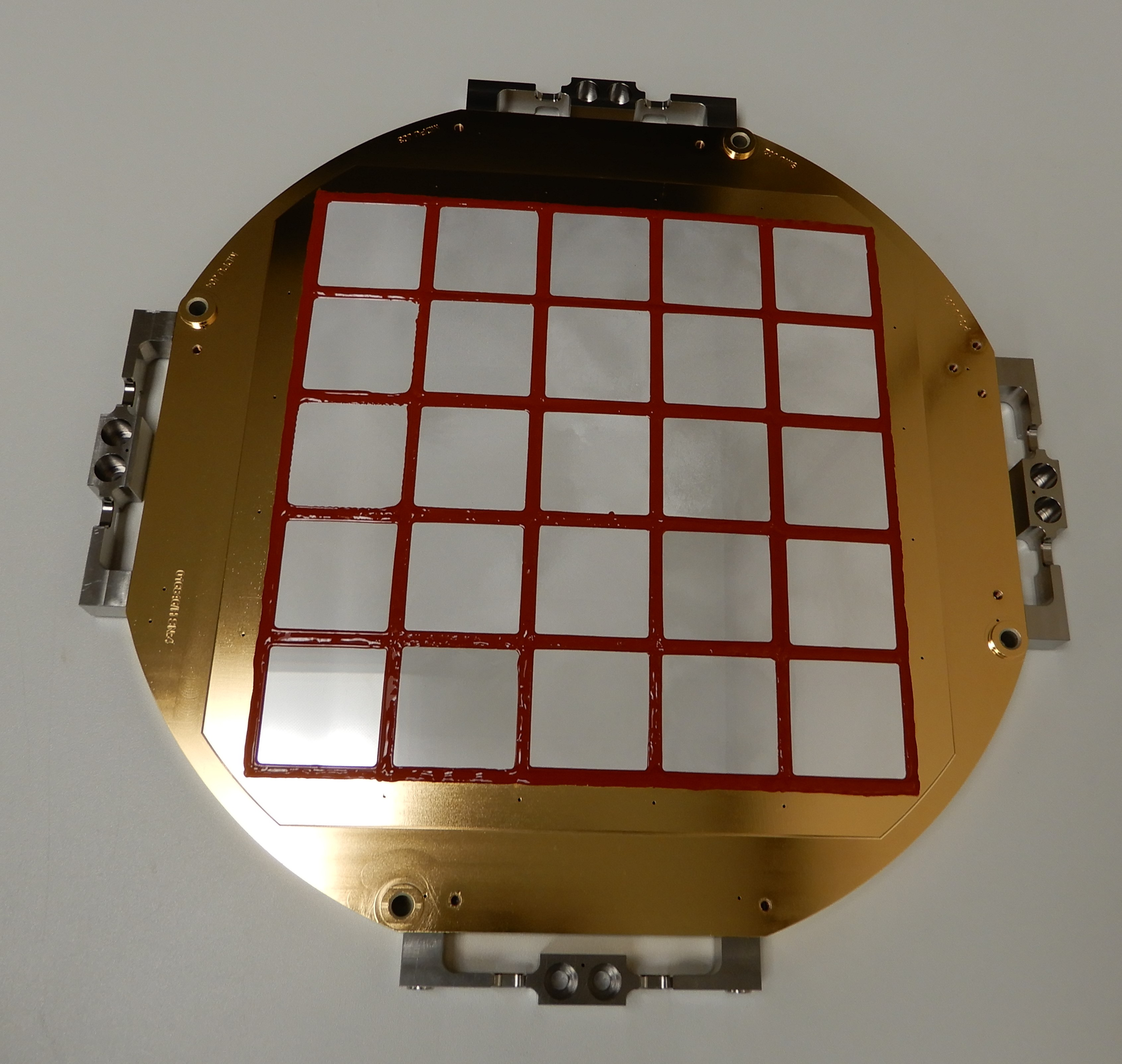}
			\caption{The flight model SVOM MXT MOP.}
	\label{fig1}
\end{figure}

The flight model (FM) MOP was constructed by bonding 25 individual MPOs onto a precision-machined spherical aluminum frame using a silicone-based adhesive. Each MPO measures 40 mm × 40 mm, with 2 mm spacing between adjacent tiles. The total mass of the fully assembled optic, including heaters, harnesses, and mounting flexures, is 1.43 kg. The completed FM MOP is shown in Figure \ref{fig1}.

The proto-flight model (PFM) MXT instrument was integrated at CNES in Toulouse and consists of the FM MOP, the PFM MXT pnCCD camera (Moita et al. this issue), thermal and mechanical ground support equipment (GSE), and a data processing unit.

\subsection{Analysis and modeling software}
\label{soft}
The data analysis was conducted using the sequential ray-tracing software Q, developed by Prof. Richard Willingale(\cite{qsoft}). This code enables detailed modeling of individual MPO responses across a range of energies and incident angles, incorporating physical characteristics and optical distortions(\cite{dickspie}). By inputting the measured properties of each MPO—such as radius of curvature (RoC), open fraction, and frame alignment parameters (including metrology-derived bias angles and distortions)—the software generates an accurate simulation of the full optic assembly.

\begin{figure}
	\centering
		\includegraphics[width=0.27\textwidth]{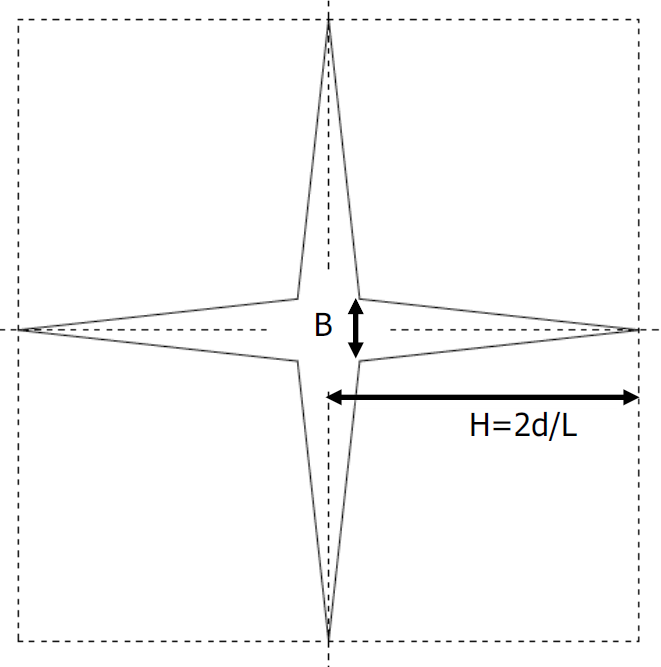}
			\caption{The lobster eye beam.}
	\label{lobeyebeam}
\end{figure}

The same software was employed to analyse MOP performance data obtained both at the PANTER X-ray test facility and in orbit. Due to the unique PSF structure produced by lobster-eye optics, conventional beam definitions (e.g., circular or square apertures) are inadequate. Instead, a lobster-eye beam method is used, defined by a four-pointed star geometry (Figure \ref{lobeyebeam}). The central core dimensions, denoted $B$, are optimized such that the beam encloses 50\% of the total detected source flux while minimizing the enclosed sky area. This yields a robust metric of angular resolution, referred to as $B_\mathrm{HEW}$.

\begin{equation}
    \begin{split}f_1(x)=1/(1+(2x/G)^2)\\\\ f_2(x)=\eta(G/H)(1-(x/H)^2)\end{split}
    \label{eq1}
\end{equation}\\
\begin{equation}
    \begin{split}f_1(y)=1/(1+(2y/G)^2) \\\\ f_2(y)=\eta(G/H)(1-(y/H)^2)\end{split}
    \label{eq2}
\end{equation}\\
\begin{equation}   
    F(x,y)=(f_1(x)+f_2(x))(f_1(y)+f_2(y))/(1+\eta)^2
    \label{eq3}
\end{equation}\\

By expanding the equations \ref{eq1},\ref{eq2} and \ref{eq3}, it is possible to calculate $G$, the Lorentzian width (FWHM), where $H=2d/L$, the collimation angle of the pores and $\eta$ is the brightness of the cross arms with respect to the central spot.

\section{On-ground calibration summary}           
\label{sect:ogcal}

Calibration of the PFM MXT instrument was conducted at the PANTER X-ray test facility (MPE) between 13$^\mathrm{th}$ of October and 19$^\mathrm{th}$ of November 2021. The campaign characterized the performance of the full PFM MOP across multiple X-ray energies. Key objectives included verification of the optic’s focal length, measurement of the effective area over the instrument's operational energy band, and assessment of off-axis response and vignetting effects.

\subsection{PANTER Calibration Campaign}
\label{pant}
The PANTER X-ray facility carries out testing of optics, optical assemblies, and full telescopes under X-ray illumination in a 130 m long X-ray beam line, using a multi-target X-ray source, capable of generating energies from 0.28 to 10.0 keV and also has a laser alignment system. At the detector end is a 3.5 m diameter, 12 m long vacuum chamber, multiple manipulator stages, and two mechanically, vibration isolated optical benches.

\begin{figure}
	\centering
		\includegraphics[width=0.4\textwidth]{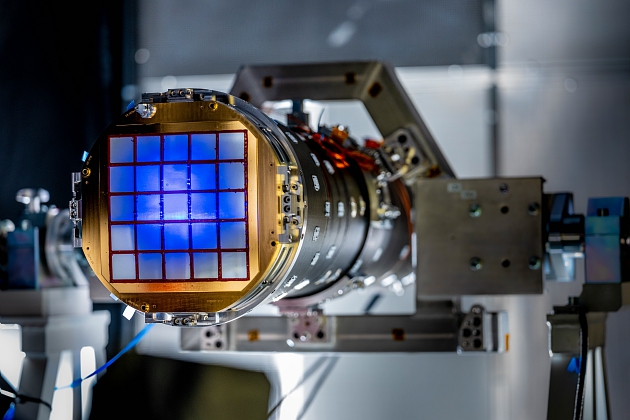}
			\caption{The PFM MXT mounted within the PANTER X-ray beam line. (Credit: CNES)}
	\label{mxtatpant}
\end{figure}

Figure \ref{mxtatpant} shows the complete PFM MXT mounted within the PANTER beam line during laser alignment. Full details of the ground calibration can be found in \cite{svomspie22}.

\begin{figure}
	\centering
		\includegraphics[width=0.3\textwidth]{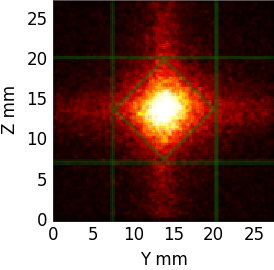}
		\includegraphics[width=0.3\textwidth]{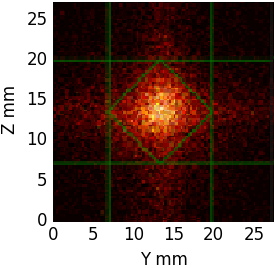}
			\caption{The X-ray image at Al-K (1.49 keV) of \textit{top:} the FM MOP at PANTER and \textit{bottom:} the full MXT at PANTER. The green lines denote areas of the image used by the analysis software.}
	\label{alkTTF}
\end{figure}

The MOP underwent two calibration phases at PANTER: first as a standalone optic, and subsequently as part of the integrated PFM MXT. On-axis images were acquired at Al-K (1.49 keV) at both the best focus and the MXT's fixed focal length; representative images are shown in Figure \ref{alkTTF}.

During standalone MOP calibration, the FWHM at Al-K was measured to be 11.23 arcmin with a focal length of 1137 mm. Based on this result and the focal length–FWHM curve derived from the standalone calibration, a prediction was made for the MXT's on-axis performance, accounting for the finite source distance used at PANTER.

Following integration into the PFM MXT, the focal length was remeasured and confirmed to be 1137 mm—consistent with the standalone result and corrected for finite source geometry. At this focal length, the FWHM was measured to be 11.75 $\pm$ 0.1 arcmin.

\subsection{VIGNETTING}
\label{vin}
The vignetting and consistency of the FWHM across the field of view was verified by rotating the full telescope bench and moving the focus to the edges of the MXT detector. The images were taken at 9 grid positions and the Al-K (1.49 keV) image is shown in the top of Figure \ref{alkinorb}. The consistency in the size and shape of the PSF is demonstrated in the 9 images shown in the top of Figure \ref{alkinorb} and the values in the central column of Table \ref{inorbres}.

\section{In-orbit calibration}           
\label{sect:iocal}

\begin{figure}
	\centering
		\includegraphics[width=0.45\textwidth]{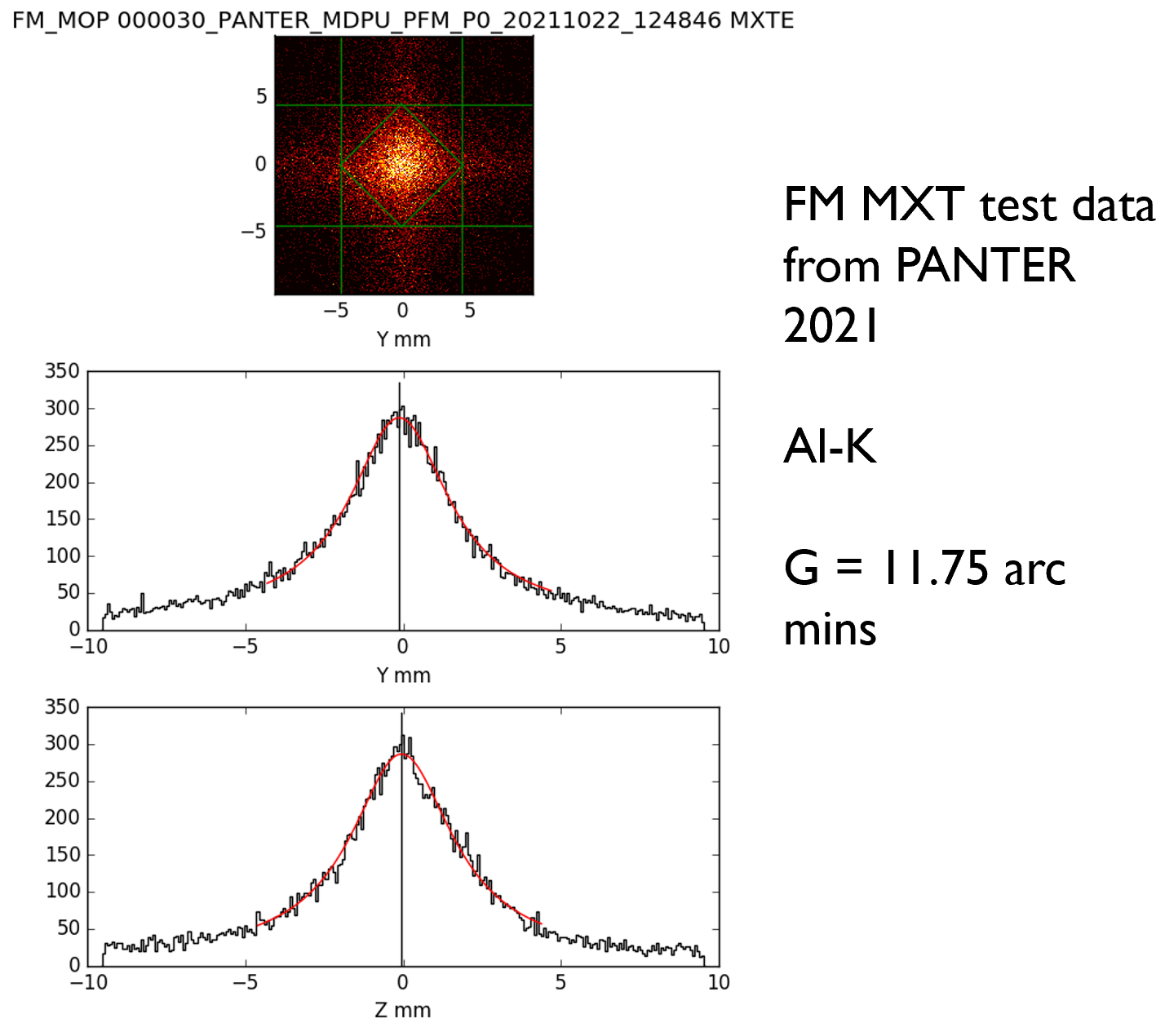}
        \includegraphics[width=0.45\textwidth]{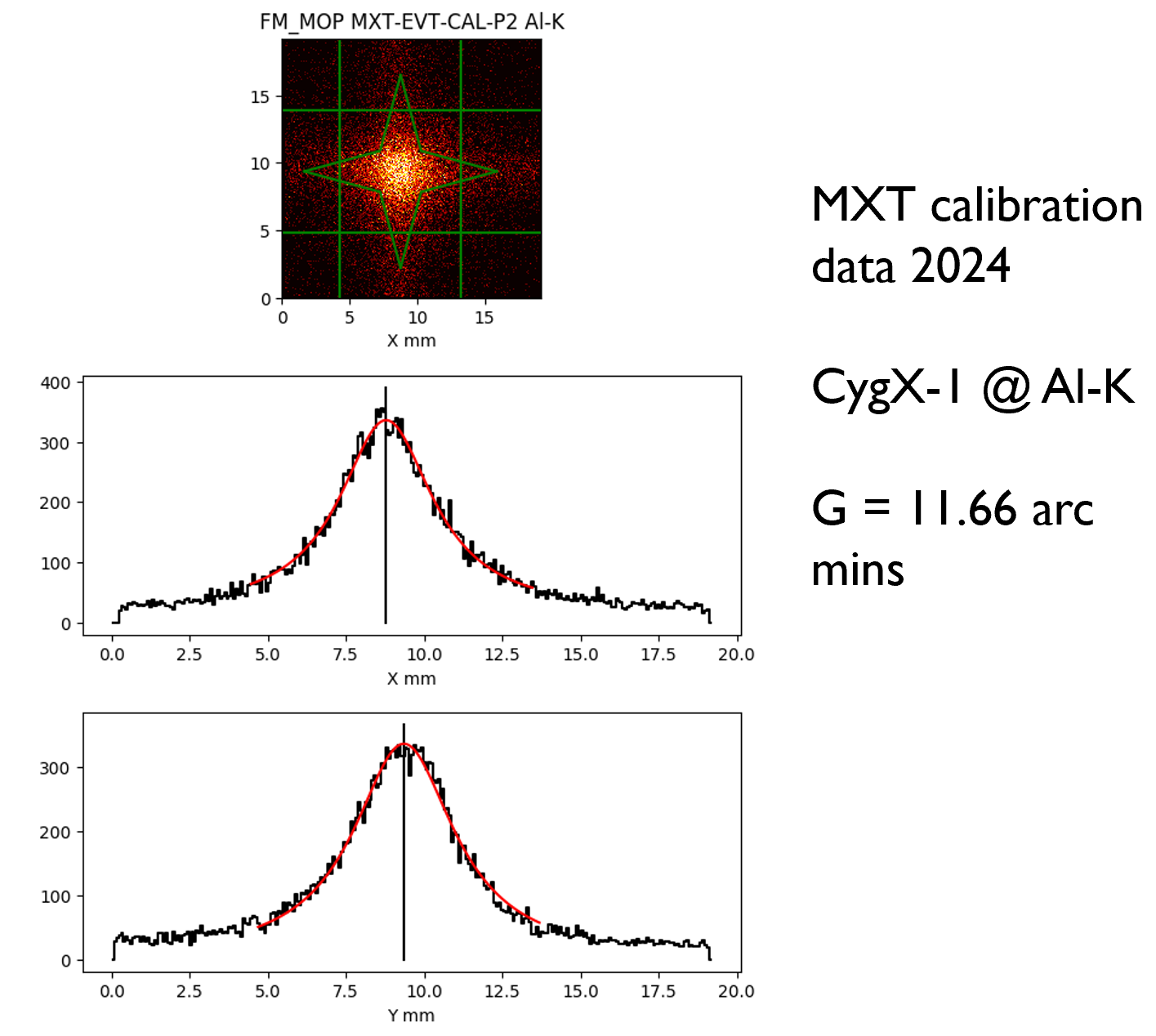}
			\caption{On-axis ($P_0$) image of an X-ray source at Al-K, both during \textit{top} on-ground calibration at PANTER and \textit{bottom} using Cyg X-1 for in-orbit calibration.}
	\label{inorbrespng}
\end{figure}

Following launch, the MXT was pointed at Cyg X-1, during the SVOM Performance Verification phase, to conduct the in-orbit calibration of the MXT optic. An on-axis image of Cyg X-1 was acquired, and Al-K photons (1.49 keV $\pm$10 eV) were isolated for analysis. The FWHM of the resulting PSF was measured and compared to the corresponding Al-K on-axis image obtained during ground calibration. The comparison is shown in Figure \ref{inorbrespng}.

A repetition of the vignetting grid measurement (Section \ref{vin}) was performed to enable direct comparison with the corresponding on-ground calibration data. Only photons at Al-K energy (1.49 keV) were selected, and images were generated for each of the nine grid positions. The grid positions are labeled with $P_0$ being in the centre, and the remaining 8 positions numbered sequentially, from left to right starting at the top left corner.

\begin{figure}
	\centering
		\includegraphics[width=0.31\textwidth]{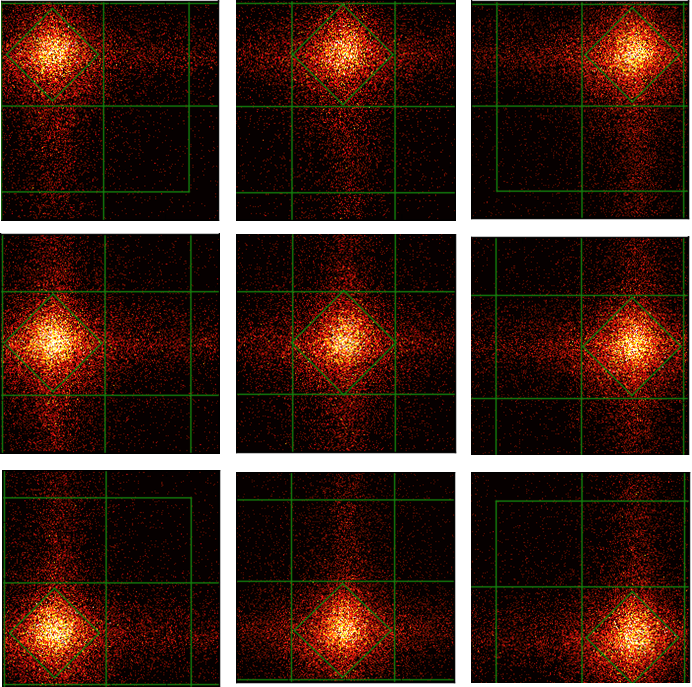}
		\includegraphics[width=0.3\textwidth]{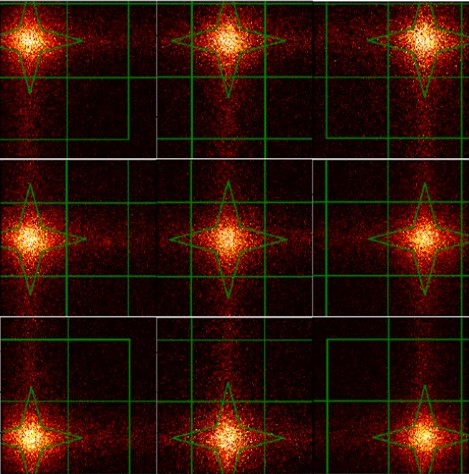}
			\caption{Vignetting grids at Al-K for \textit{top} on-ground calibration and \textit{bottom} in-orbit calibration. The green lines denote the lobster eye beam regions used to measure the FWHM.}
	\label{alkinorb}
\end{figure}

The lobster-eye beam method was applied to determine the full width at half maximum (FWHM) at each position, both in orbit and on the ground. The resulting in-orbit image is shown on the right in Figure \ref{alkinorb}, and the corresponding FWHM values are listed in Table \ref{inorbres}. The differences in the majority of cases is due to statistics, however, the in-orbit $P_2$ (top centre) and $P_8$ (bottom right) images show a much cleaner and brighter double reflection spot then on the ground. This is likely due to the active MPOs in that area being in focus for an infinite source, whilst they were out of focus within the finite beamline thus broadening the PSF.

\begin{table}
    \centering
    \begin{tabular}{c|c|c}
        \textbf{Source position} & \textbf{On-ground} & \textbf{In-orbit}\\
        \hline
        P0 & 11.75 $\pm$ 0.18 & 11.66 $\pm$ 0.17\\
        P1 & 11.12 $\pm$ 0.16 & 11.91 $\pm$ 0.18\\
        P2 & 13.00 $\pm$ 0.20 & 11.21 $\pm$ 0.18\\
        P3 & 11.85 $\pm$ 0.18 & 11.78 $\pm$ 0.18\\
        P4 & 11.47 $\pm$ 0.17 & 12.08 $\pm$ 0.18\\
        P5 & 12.31 $\pm$ 0.18 & 11.28 $\pm$ 0.16\\
        P6 & 11.56 $\pm$ 0.17 & 11.81 $\pm$ 0.18\\
        P7 & 12.14 $\pm$ 0.18 & 11.64 $\pm$ 0.18\\
        P8 & 12.30 $\pm$ 0.18 & 10.88 $\pm$ 0.16\\
    \end{tabular}
    \caption{Comparison of the lobster eye beam FWHM measurements, at each of the 9 vignetting grid positions, both during on-ground calibration and in-orbit calibration.}
    \label{inorbres}
\end{table}

\section{Electron Diverter Design and Performance}
\label{sect:ed}
The MPOs onboard MXT focus X-rays via grazing incidence reflection along square microchannels. This geometry, while effective for X-ray focusing, also provides a direct line of sight from the focal plane to the surrounding environment, allowing high-energy charged particles — particularly electrons and protons — to reach the detector.

Electrons in the 0.01–10 MeV range can deposit significant energy in silicon-based detectors such as the MXT pnCCD, increasing background levels and reducing sensitivity. To mitigate this, a system of permanent magnets was installed behind the optics to deflect incident electrons away from the focal plane. The electron diverter was mounted on a separate frame and positioned as close as possible to the rear of the optic to maximize deflection efficiency while minimizing vignetting and avoiding mechanical interference during launch and testing.

The diverter design was completed using an electron ray-tracing model originally developed by one of us (Richard Willingale, University of Leicester, \cite{qsoft}), previously used in the design of diverters for ROSAT WFC, XMM-Newton and Swift. The magnetic field is modeled as a superposition of discrete dipoles in a user-defined configuration. Each physical magnet is approximated as a pair of dipoles separated by 3 mm along the optical axis — corresponding to the physical length of the magnets. To preserve magnetic strength, each modeled dipole carries half the moment of the full magnet. 

Electron trajectories were initialized at random positions on the optic’s rear plane, directed toward the detector center to simulate perfect mirror focusing. Angular deviations can be added to emulate scattering if required. Electrons are propagated through the magnetic field until they either (1) reflect through the optic plane, (2) impact the inner telescope tube wall (modeled as a cylinder), (3) reach the focal plane, or (4) are trapped in the field for over 10,000 iterations, corresponding to small gyroradii electrons confined by the diverter.

\begin{figure}
	\centering
		\includegraphics[width=0.3\textwidth]{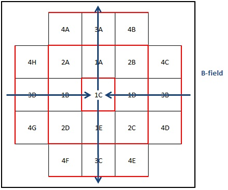}
        \includegraphics[width=0.25\textwidth]{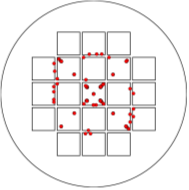}
			\caption{The MXT electron diverter \textit{top} the design showing the direction of the magnetic dipoles and \textit{bottom} the map of impacts on the focal plane, showing the electron diverter residual transmission.}
	\label{eddesign}
\end{figure}

The selected configuration, shown in Figure \ref{eddesign} (top), exhibited the lowest focal plane impact rate across the widest range of electron energies. However, due to the central MPO, a direct path remained through which high-energy electrons could reach the detector. This residual transmission is visible in the simulated impact map (Figure \ref{eddesign}, right) and could not be eliminated without obstructing the MPO and reducing the effective area of the telescope. The electron diverter design is shown with the qualification model optic design, which only had 21 MPOs. For the FM optic, the magnet positions and orientations were extended to include the corner MPOs. Similar electron diverters have been designed for other lobster eye missions(\cite{valed}).

\begin{figure}
	\centering
		\includegraphics[width=0.35\textwidth]{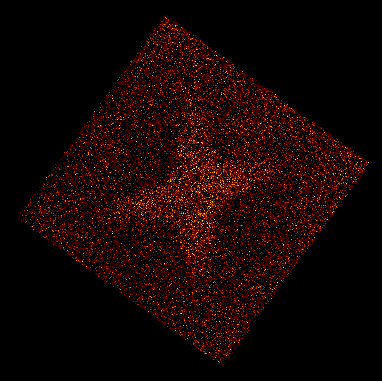}
			\caption{An example of an in-orbit background image showing typical patterns caused by high energy electron residual transmission. Image taken on March 25$^\mathrm{th}$ 2025 (OBSID 1996493423) with an exposure time of 2.2 ks over the 0.3 - 10 keV band and displayed with a linear scale.}
	\label{inorbleak}
\end{figure}

In orbit, this residual transmission of electrons produced unexpected features in background images (Figure \ref{inorbleak}). Comparison with the simulation confirmed the pattern matched the predicted impact distribution of the transmission. As a result, the contamination is now treated as an instrumental background component and will be removed during image processing.

\section{Conclusions}
\label{sect:conclusion}
The MXT telescope, employing a lobster eye optic based on 25 slumped MPOs, has been successfully calibrated both on-ground and in orbit. The design, optimized for a $\sim$1$^{\circ}$ detector-limited field of view, maintains a uniform point spread function (PSF) across a $\sim$6$^{\circ}$, leveraging the inherent wide-field advantages of lobster eye geometry. On-ground calibration, conducted at the PANTER X-ray test facility, established an instrument focal length of 1137 mm and an on-axis FWHM of 11.75 arcmin at Al-K (1.49 keV). In-orbit calibration using Cyg X-1 demonstrated strong consistency with these ground-based measurements, validating the stability and performance of the optic post-launch.

In addition to X-ray performance, the MXT design includes a magnetic electron diverter intended to reduce charged particle background at the detector. Using a dedicated electron ray-tracing model, the optimal dipole configuration was identified to minimize focal plane electron impacts while preserving optical throughput. However, due to the direct line-of-sight through the central MPO, a residual electron transmission remains, visible both in simulations and in-orbit data. This contamination manifests as a distinct features in background images. While the residual transmission could not be physically removed without vignetting the optics, its characteristic signature enables effective removal through post-processing.

\begin{acknowledgements}
The Space-based multi-band astronomical Variable Objects Monitor (SVOM) is a joint Chinese-French mission led by the Chinese National Space Administration (CNSA), the French Space Agency (CNES), and the Chinese Academy of Sciences (CAS). We gratefully acknowledge the unwavering support of NSSC, IAMCAS, XIOPM, NAOC, IHEP, CNES, CEA, and CNRS.\\
This research used the ALICE High Performance Computing Facility at the University of Leicester.
\end{acknowledgements}



\label{lastpage}

\bibliographystyle{raa}
\bibliography{bibtex}

\end{document}